\documentclass[%
,twocolumn%
,secnumarabic%
,tightenlines%
,amssymb, amsmath,nobibnotes, aps,nofootinbib,showpacs,]{revtex4}
\usepackage{epsfig}%
\usepackage{graphicx}%
\expandafter\ifx\csname package@font\endcsname\relax\else
 \expandafter\expandafter
 \expandafter\usepackage
 \expandafter\expandafter
 \expandafter{\csname package@font\endcsname}%
\fi

\begin{document}
  \title{The emission of Gamma Ray Bursts  as a test-bed  for  modified gravity}
\author{S. Capozziello$^{a,b,c}$, and G. Lambiase$^{d,e}$}
\affiliation{$^a$Dipartimento di  Fisica, Universit\'a di Napoli "Federico II", Via Cinthia, I-80126 - Napoli, Italy.}
\affiliation{$^b$INFN Sez. di Napoli, Compl. Univ. di Monte S. Angelo, Edificio G, Via Cinthia, I-80126 - Napoli, Italy,}
\affiliation{$^c$Gran Sasso Science Institute (INFN), Via F. Crispi 7, I-67100, L'Aquila, Italy,}
\affiliation{$^d$Dipartimento di Fisica "E.R. Caianiello", Universit\'a di Salerno, I-84084, Fisciano (Sa), Italy,}
\affiliation{$^e$INFN - Gruppo Collegato di Salerno, Italy.}
\date{\today}

\begin{abstract}
The extreme physical conditions of Gamma Ray Bursts  can constitute a useful observational laboratory to test  theories of gravity where very high curvature regimes are involved.
Here we propose a sort of {\it curvature engine} capable, in principle, of explaining the huge energy emission of Gamma Ray Bursts.
Specifically, we investigate the emission of radiation by charged particles non-minimally coupled to the gravitational background where higher order curvature invariants are present.
The coupling gives rise to  an additional force  inducing a non-geodesics motion of particles. This fact  allows
 a strong  emission of radiation by {\it gravitationally accelerated} particles. As we will show with some specific model, the energy emission is
of the same order of magnitude of that  characterizing the Gamma Ray Burst physics.  Alternatively, strong curvature regimes   can be considered  as a natural mechanism for the generation of
highly energetic astrophysical events. Possible applications to cosmology are discussed.
\end{abstract}
\pacs{04.50.+h, 04.50.Kd,  98.70.Rz, }

\maketitle

General Relativity can be considered one of the major achievement of human thought. Such a theory deals with space and time as dynamical variables and introduces new concepts as  black holes, cosmic expansion, time travels and so on. However, it has not been deeply investigated at the ultraviolet  regime where strong deviations from the standard Hilbert-Einstein picture could emerge \cite{review1}.  On the other hand, the Einstein theory seems to require a lot of new ingredients in order to fit completely the gravitational dynamics as dark energy and dark matter at infrared scales and some new approach in order to deal with gravitational interaction at fundamental level (quantum gravity).

This state of art have to be revised in view of a unitary theory encompassing the gravitational phenomenology at all scales. In this perspective, several alternative or modified theories of gravity have been proposed first of all to address the  shortcomings related to the Cosmological Standard Model. In particular,  taking into account higher oder curvature invariants than the simple Ricci scalar $R$ naturally gives rise to inflation that removes the primordial singularity and explain the so called flatness and horizon problems \cite{starobinsky}. The fundamental physics motivation of this (approach) and the related ones is that, at high curvature regimes, further curvature invariants have to be considered in order to construct self-consistent effective actions in curved spacetime \cite{birrell}. In some sense, the high curvature regimes requires the introduction of further invariants. This is not the final step towards the quantum gravity, but it is an effective picture that works well at least at one-loop level \cite{barth}.

The lesson is that towards ultraviolet regimes (or in high density regimes) the General Relativity has to be improved adding further curvature corrections. Such an approach seems to work also for very compact objects as  massive neutron stars. As recently shown in \cite{tov,upasana}, higher curvature terms can naturally provide a mechanism to improve the masses of compact objects without invoking exotic forms of the equation of state. In some sense, the curvature acts as an engine that give rise to a further pressure term in the Tolman-Oppenheimer-Volkov equation describing the physics of the neutron star.

A similar mechanism could work also for  one of the  most intriguing astrophysical phenomena i.e.  the Gamma Ray Bursts (GRBs).
They are short and intense   pulses of  $\gamma$ rays
discovered in  1973 \cite{GRBsdiscovery}.  They are, without any doubt,  the most energetic events in the Universe a part the Big Bang.
GRBs are essentially characterized by the following properties (for a reviews, see \cite{review}):
 \begin{itemize}
   \item They arrive from cosmological distances from random directions in the sky.
   \item Their have a extraordinary large energy outputs of the order
   \begin{eqnarray}\label{WGRBs}
     W_{\rm{GRBs}}&\sim & 10^{51}\div 10^{54} \rm{erg/s} \\
     &\sim& 4\times (10^{30}\div 10^{33})\rm{GeV}^2\,. \nonumber
   \end{eqnarray}
   \item Their spectra are non-thermal, and, as widely believed, are due to synchrotron radiation.
   \item Their duration varies  from $10^{-3}$s to $10^3$s.
 \end{itemize}
However, recent results  show that most of GRBs are narrowly beamed and the corresponding energies are
$10^{51}$ergs/s \cite{frail}, making them comparable to supernovae
in total energy release.


Although some features of GRBs must be still understood,
there is  agreement between  observations and the so called fireball model. According to the latter,
GRBs are produced via a dissipation of the kinetic energy of  ultra-relativistic
flows. In this model, the GRB itself is produced by internal dissipation within the flow, while the
afterglow\footnote{Models predict that GRBs are followed by a lower-energy afterglow and
in some cases,   radio afterglows have been observed several years after the bursts.},
a long-lasting emission in the x-ray, optical, and radio
wavelengths, is produced via external shocks with the medium.

Owing to the several  observations of GRBs and of their afterglows, it has been possible to constrain the fireball model
that describes the emitting regions. There is, however, no direct evidence about the inner engine able to generate GRBs
and produce the ultrarelativistic flow. For a  physical  characterization of  these phenomena, in particular
the energetic requirements and the time scales, one deduces that GRBs are correlated with the formation of black holes
(via a stellar collapse) or a neutron star merger.
Moreover, the requirement of the  long  activity of the inner engine in the fireball model (typically greater than 10 s)
suggests an inner engine built on an accreting black hole. This agrees with the the fact that GRBs are associated with star
forming regions,  indicating that GRB progenitors are massive stars. Finally, the appearance of supernova
bumps in the afterglow light curve (most notably in GRB 030329 \cite{GRB}) suggests a correlation of GRBs with supernovae
and stellar collapse.

A part the careful description of the emission in the fireball model, the very final origin of such a strong energetic mechanism is far to be fully understood at fundamental level.
According to the above considerations, high curvature regimes could play an important role in this framework.
Here we want to  discuss the emission of radiation by
charged particles nonminimally coupled to gravitational background showing that curvature, coupled with matter, could be a natural engine for GRB emission.  In \cite{obhukov},
it is  shown that matter, nonminimally coupled to the background, may induce an additional pressure-like term
in the equation of motion, so that particles do not move along geodesic (leading, in some way, to a violation of the
equivalence principle). This could provide a  mechanism for the production of GRBs with the observed energy releases  (\ref{WGRBs}).
In some sense, the curvature, generating an extra pressure term,  plays the same role as in the neutron stars giving rise to  a change of the mass - radius relation $M - {\cal R}$ \cite{tov}.
Here we shall neglect some physical effects as the quantum particle
(pairs) creation induced by the high space-time curvature or the
interaction of charged particles with magnetic fields, and, instead of using Klein-Gordon or Dirac equation for 
studying the evolution of particles in the gravitational field, we shall confine ourselves to a classical description.
Besides, also the back-reaction effects will be neglected in our analysis
being only interested to the emission of radiation induced by the
acceleration.  

It is well known that particles with high acceleration generate enormous streams of photons by {\it brehmsstrahlung}.
The power radiated away by a particle of charge $q$ is estimated to be (for convenience, we shall use both mks and
natural units) \cite{landau}
 \begin{equation}\label{Wnoncov}
 W=-\frac{2q^2}{3}\,\vert {\ddot {\bf x}}\vert^2\,,
 \end{equation}
where $\vert {\ddot {\bf x}}\vert \equiv |{\bf a}|$ is the modulus of the acceleration. However, in a curved spacetimes, to which we are mainly interested, the above equation generalizes to
  \begin{equation}\label{Wcov}
 W=-\frac{2q^2}{3}\,\vert {{\cal D}^2 x^\alpha}\vert^2\,,
 \end{equation}
where now ${\cal D}^2 x^\alpha$ represents the covariant four-acceleration of particles. Clearly for particles moving along a geodesic, the four-acceleration is zero, and no radiation can be emitted. This is not the case in models recently proposed  \cite{obhukov}.
Let us assume  the total interaction Lagrangian  given by
 \begin{equation}\label{Ltot}
   {\cal L}= {\cal L}_{grav}+F{\cal L}_{mat}\,,
 \end{equation}
where ${\cal L}_{grav}$ and $F$ can arbitrarily depend on the spacetimes metric and curvature tensors according to the effective interaction considered.
The last term in (\ref{Ltot}) describes theories with nonminimal coupling between matter and functions depending on curvature invariants.
In general, one may consider the possibility that the coupling $F$ is a more involved function depending on nine parity-even invariants \cite{obhukov}, so that $F=F(i_1, \ldots, i_9)$, where $i_1, \ldots, i_9$ are the scalar curvature invariants constructed by means of Ricci and Riemann tensors
 \begin{equation}\label{inv123}
 i_1 \equiv R^2,  \quad i_2 \equiv R_{\mu\nu}R^{\mu\nu}, \quad
 i_3 \equiv R_{\mu\nu\alpha\beta}R^{\mu\nu\alpha\beta}
 \end{equation}
  \[
  i_4 \equiv R_{\mu\nu}^{\phantom{\mu\nu}\alpha\beta}
  R_{\alpha\beta}^{\phantom{\alpha\beta}\sigma\rho}
  R_{\sigma\rho}^{\phantom{\sigma\rho}\mu\nu}\,, \quad
  i_5\equiv R^\mu_{\phantom{\mu}\nu}R^\nu_{\phantom{\nu}\rho}
  R^\rho_{\phantom{\rho}\sigma}\,,
  \]
  \[
  i_6\equiv R^\mu_{\phantom{\mu}\nu}R^\nu_{\phantom{\nu}\rho}
  R^\rho_{\phantom{\rho}\sigma}R^\sigma_{\phantom{\sigma}\delta}\,,
  i_7\equiv R^{\mu\nu}D_{\mu\nu}\,,
  \]
  \[
  i_8\equiv D_{\mu\nu}D^{\mu\nu}\,,\quad i_9 \equiv D_{\mu\nu}D^{\nu\rho}R^\mu_{\phantom{\mu}\rho}\,,\quad
  D_{\mu\nu} \equiv R_{\mu\nu\rho\sigma}R^{\nu\sigma}\,.
  \]
  In general, also the  Gauss-Bonnet topological invariant
  \begin{equation}
  {\cal G}\equiv R^2-4R_{\mu\nu}R^{\mu\nu}+R_{\alpha\beta\mu\nu}R^{\alpha\beta\mu\nu}
  \end{equation}
  can be considered a term playing a important role in the matter-gravity interaction \cite{barth, felix}.
The equations of motion for (extended) test bodies are derived from the energy-momentum conservation law. More specifically, by using the Synge expansion technique \cite{synge} and covariant multipolar approximation scheme \cite{approx}, it turns out that four-acceleration is given by
 \begin{equation}\label{4acc}
   {\cal D}^2 x^\alpha\equiv {\dot v}^{\alpha}=\frac{\xi}{m}\left(\delta^\alpha_\beta
   -v^\alpha v_\beta\right) \nabla^\beta A\,.
 \end{equation}
In this equation, $A$ is related to the function $F$ as
 \begin{equation}\label{AlnF}
 A=\ln F\,,
 \end{equation}
$m$ is the mass of test particle and $v^\alpha$ its velocity, and finally,  the constant $\xi$ is a quantity that depends on the matter distribution
 \begin{equation}
 \xi = \int_{\Sigma(s)} {\cal L}_{mat} w^{x_2} d\Sigma_{x_2}\,,
 \end{equation}
where $s$ is the proper time of the particle and the integration is over a spatial hypersurface (in general $\xi$ does depend on gravitational background and
parameters characterizing the test particle; however, in a multipolar approximation, $\xi$ corresponds to a free particle \cite{obhukov,iorio}).
Eq. (\ref{4acc}) therefore implies that a massive particle moves
non-geodesically along its world-line owing to the presence of the additional {\it force} generated by the non-minimal coupling curvature function $F$. Consequences of (\ref{4acc}) have been recently studied in cosmology \cite{ishak} and celestial mechanics \cite{iorio}. Stringent constraints on $|\xi \nabla_\alpha A|$ are provided by COBE and GP-B satellites,
$|\xi \nabla_0 A|\simeq 2\times 10^{-4}\rm{kg/sec}=7.4\times 10^{-2}\rm{GeV}^2$ and $|\xi \nabla_i A|\simeq 2\times 10^{-10}\rm{g/sec}=7.4\times 10^{-8}\rm{GeV}^2$ \cite{iorio}.

Squaring (\ref{4acc}) and using the normalization condition $v^\alpha v_\alpha=-1$, one obtains that the emitted radiation (\ref{Wcov}) reads
 \begin{equation}\label{Wemitted}
   W=-\frac{2}{3}\frac{q^2}{m^2}\xi^2 K^2\,,
 \end{equation}
 where
 \begin{equation}\label{Kdef}
   K^2=|\nabla_\alpha A|^2-(v^\alpha \nabla_\alpha A)^2\,.
 \end{equation}
 It is interesting to note that our model could be considered as the counterpart of the GRBs emission mechanism due to the synchrotron radiation. In the last case, in fact, using the equation of motion for a charged particle in a magnetic field ${\bf B}$, one finds ${\displaystyle W_{syn}=\frac{4}{3}\sigma_{Th} U_B \gamma^2}$ \cite{upasana}, where $U_B=B^2/8\pi$ is the energy density of the magnetic field, $\gamma$ the product of the relativistic factors of the shocked fluid and electron, and  ${\displaystyle \sigma_{Th}=\frac{8\pi}{3}r_s^2\sim 6.6 \times 10^{-29}\text{m}^2\sim 1.6 \times 10^3\text{GeV}^{-2}}$ the Thompson cross section. The magnetic field varies strongly from the inner engine $10$Km, $B\sim 10^{15}$G, to the internal shock $10^8-10^{10}$Km, $B\sim 10^6$G, to  the afterglow $10^{11}-10^{13}$Km, $B\sim 1$G.
To get the typical GRBs emission, one  needs high values of the relativistic factor $\gamma$ that is dependent on the emitting regions.
Of course both mechanisms, the conventional synchrotron mechanism and our mechanism, based on the non-minimal gravitational coupling, are efficient for the generation of the observed GRBs energies, provided that the parameters are properly chosen. However, besides the velocity of charged particles, the {\it gravitational} mechanism does depend on $\xi |\nabla_\alpha A|$ (see Eqs. 
(\ref{AlnF}), (\ref{Wemitted}), and (\ref{Kdef})),
which means that a rapid variation of the gravitational coupling $F$ (in time or space), or a large $\xi$  can induce an enhancement of the emitted radiation.

\begin{figure}
  \centering
  \includegraphics[width=8cm]{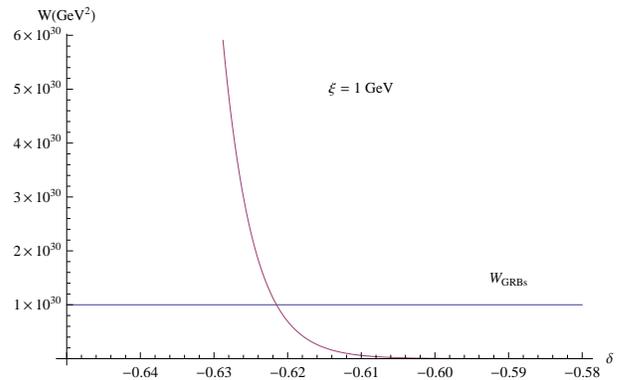}\\
  \caption{Plot of  Eq. (\ref{Wemittedexp}). The used values  are $\Gamma=10$, $\xi=1$, and $m=0.5$eV.}
  \label{Fig1}
\end{figure}

In any case, a full analysis of (\ref{Wemitted}) requires to fix the form of $A$, hence of $F$. In the following we consider some specific case:

\begin{enumerate}
  \item Let us first consider the case where the variation of the function $K^2$ is negligible over the time scale of GRBs duration. This means to consider $K$ constant (i.e. $A\sim C_\mu x^\mu$, where $C_\mu$ is a
      constant four-vector). In order to achieve  the emitted power $10^{51}-10^{54}$erg/sec, the parameter $\xi$
      has to be  constrained to
      \begin{eqnarray}\label{Kgeneric}
      \xi K &\lesssim &(10^{10}-10^{11.5}) \rm{kg/sec} \\
       &=&4.1\times (10^{12}-10^{13.5})\rm{GeV}^2\,. \nonumber
      \end{eqnarray}
  \item Consider now the case where the background is described by a Schwarzschild geometry. Outside the gravitational source,  the Ricci tensor vanishes, while the Riemann tensor does not. In such a case, one can assume that the coupling $F$ is only a function of the invariant $i_3=\displaystyle{12\frac{r_s^2}{r^6}}$, where $r_s=2GM$ is the Schwarzschild radius of the gravitational mass. As a  more general form,  we can choose
      \begin{equation}\label{Fldelta}
      F(i_3)=(\lambda^4 i_3)^\delta\,,
      \end{equation}
      with $\lambda$ a constant of dimensions $[\rm{length}]$ (or $[\rm{energy}]^{-1}$) and $\delta$ a dimensionless  constant. The functions $F$ and $A$ do only depend on the radial variable $r$. For the sake of simplicity,  we also assume that the motion of the particle is radial ($v^\alpha = (v^0, v^r, 0, 0)$. Eq. (\ref{Kdef}) gives
      $K^2=\displaystyle{\frac{36\delta^2}{\Gamma^2 r^2}}$, with $\Gamma=[1-(v^r)^2]^{-1/2}$. Referring to an electron particle, with $m=0.5$MeV and $q=e=2.8\times 10^{-1}$, and astrophysical objects with characteristic Schwarzschild radius $r_s = 10$km (the mass of black hole are typically of the order $(3\div 10)M_\odot$, where $M_\odot$ is the solar mass), we get that the emitted power is
      \begin{eqnarray}\label{WemittedSch}
        W &=& 24 \frac{e^2}{m^2\Gamma^2 r^2}\,(\xi \delta)^2 \\
         &\simeq & 4\times 10^{-33}
         \frac{(\xi \delta)^2}{\Gamma^2}\left(\frac{0.5\rm{MeV}}{m}\frac{10\rm{km}}{r}\right)^2\,. \nonumber
      \end{eqnarray}
      For $\Gamma=10$ one gets $W\sim 10^{51}$erg/sec $\sim 4\times 10^{30}$GeV, Eq. (\ref{WGRBs}) provided $\xi\delta \lesssim 10^{33}$GeV.

  \item Finally, consider the following form of $F$:
   \[
   F(i_3) = e^{(\lambda^4 i_3)^\delta}\,.
   \]
For a Schwarzschild background,  one finds that the emitted power is given by
   \begin{eqnarray}\nonumber
     W  &=& 24 \times 12^{2\delta}10^{-32}
     \left(\frac{0.5\rm{MeV}}{m}\frac{10\rm{km}}{r_s}\right)^2 \frac{e^2}{\Gamma^2}\times \\
     & & \times \left(\frac{\delta \xi}{\rm{GeV}}\right)^2  \left(\frac{\lambda}{r_s}\right)^{8\delta}
     \left(\frac{r_s}{r}\right)^{12\delta}\rm{GeV}^2\,. \label{Wemittedexp}
   \end{eqnarray}
For a characteristic length $\lambda$ of the order of the Schwarzschild radius, $\lambda\sim r_s$,
and for distances $r\sim 10^{15}\rm{cm}=10^9 r_s$, corresponding to the distance where the shock produces a GRB,
one gets the emitted power (\ref{WGRBs}) provided $\xi\sim {\cal O}(1)$GeV and $\delta \simeq -0.62$. 
This is represented in Fig. \ref{Fig1}.

In the case where the background is described by a Kerr  spacetime, the invariant $i_3$ reads
      \begin{equation}\label{Fi3Kerr}
      i_3 = 12\frac{r_s^2}{r^6}I(x)\,,
      \end{equation}
   where
    \[
    I(x) \equiv \frac{(1-x^2)[(1-x^2)^2-16x^2]}{(1+x^2)^6}\,,
    \]
    with $x=ay/r$ and $y=\cos\theta$.  The constant $a$ is related to the angular momentum $J$ of the gravitational source $a=J/M$. For $x\ll 1$, i.e. $r\gg ay$, the function $I$ approaches to 1, and one recovers the Schwarzschild results. For $x\approx 1$  and using (\ref{Fi3Kerr}) and (\ref{Kdef}), one gets that the emitted energy is given by
    \begin{equation}\label{WKerr}
     W_{\rm{Kerr}}  = W \eta^{2\delta-1}\,,
   \end{equation}
   where $W$ is defined in (\ref{Wemittedexp}) and $\eta = x^2-1 \ll 1$. Therefore we find that for a Kerr geometry
   there appears an additional factor $\eta^{2\delta-1}$, whose effect is to amplify the emission power, i.e. $\gg 1$, provided $\delta< 1/2$.
\end{enumerate}

{For the Sun the metric can be assumed a Schwarzschild-like, with $r_\odot\sim 7\times 10^{5}\text{km}$. Eq. (\ref{WemittedSch}), as well as (\ref{Wemittedexp}), imply that the emitted power is $W_\odot \sim 10^{-12}W$,  so the effect is strongly suppressed and negligible at Solar System scales.

In conclusion,  we  propose a new mechanism for the  GRBs emission.
The mechanism is based on the non-minimal coupling of matter with the gravitational background, which
gives rise to an additional term in the equation of motion inducing a non-geodesic propagation of test particles. This aspect is very
important in processes where the radiation is emitted by accelerated particles. In fact in a pure General Relativity approach, one has that ${\cal D}^2 x^\alpha = {\ddot x}^\alpha +\Gamma^\alpha_{\mu\nu}{\dot x}^\mu {\dot x}^\nu=0$, and therefore charged particles cannot emit radiation. On the other hands, if non-minimal curvature coupling are properly taken into account,
 then ${\cal D}^2 x^\alpha \neq 0$, see Eq. (\ref{4acc}), and radiation can be emitted.

Although we have investigated some particular cases for  the non-minimal coupling function $F$, focusing on models where $F$ does only depend on the invariant $i_3$, our results are more general, as follows from  Eq. (\ref{Wemitted}).
Conversely,  we can say that GRBs  can be a formidable test-bed to select reliable  alternative theories of gravity since, from the above mechanism, natural constraints emerge. Furthermore, we have to point out that here we considered only classical particles. In a more refined treatment, quantum description has to be considered.

A final issue has to be discussed in detail. There are some theoretical suggestions indicating that GRBs could be used as formidable standard candles or, at least,  as distance indicators, in cosmology \cite{izzo}. The  point is that they would allow to extend the supernovae SNeIa Hubble diagram up to redshifts of the order $z\sim 9\div 10$ greatly improving the results of the today precision cosmology \cite{izzo2}. The main shortcoming of such an approach is that it is very difficult to "standardize" the GRB light curve finding out peculiar features as, for example, the Phillips slope \cite{Phillips} ruling the luminosity decreasing of SNeIa. This means that the GRB light curves appear without definite slopes and features and this fact, a part other calibration problems,  prevent to use them as standard candles. 
In this sense, GRBs are {\it not standard candles}, since they have no known and well-defined luminosity relation. Due to this lack, one has to find other approaches to use GRBs as cosmological indicators. A  solution consists  in finding correlations between their photometric and/or spectroscopic properties. In  literature there some of these relations have been pointed out  \cite{Schaefer}. For example, the relation discussed in \cite{Amati1} relates, for a given GRB,  the emitted isotropic energy   with the peak energy in the rest-frame  of the  electromagnetic spectrum. This relation has been   used to constrain some  cosmological  parameters as shown in  \cite{Amati2}. However, there is no physical link between this kind of relations and the mechanism related to  the emission of  GRBs.

The present proposal, beside being  a possible test-bed for modified gravity, could constitute, vice-versa a theoretical suggestion in this perspective. In fact, if the engine generating the emission is related to some curvature invariant, as in the cases described above,  the underlying cosmological models have to be compatible with the presence of higher order curvature terms. In particular, the accelerating behavior of the Hubble fluid should be generated by higher order gravity both at early times \cite{starobinsky} and at late time \cite{francaviglia}. In this sense, cosmological models generating accelerated expansion have to be compatible with the presence of GRBs acting as standard candles along the cosmological evolution. Specifically, the reported geometrical mechanism acting as the engine for the GRB emission could give constrains on  the cosmological models generating inflation at early time and dark energy at late time. Further and detailed investigations will be the argument of forthcoming papers. 

\vspace{0.2in}

\acknowledgments
S.C. acknowledges INFN Sez. di Napoli (iniziativa specifica TEONGRAV).  G.L.  thanks the ASI
(Agenzia Spaziale Italiana) for partial support through the contract
ASI number I/034/12/0. The authors acknowledge the constructive suggestions given by the Referee that allowed to improve the paper.

\end{document}